\documentclass[journal]{IEEEtran}
\usepackage{amsmath, amssymb, amsfonts, amsthm}

\usepackage{xcolor}
\usepackage{algorithm, algpseudocode, float, authblk, fancyhdr}
\newcommand\E{\mathbb{E}}
\newcommand\R{\mathbb{R}}

\newcommand\md{\mathrm{d}}
\newtheorem{lemma}{Lemma}
\newtheorem{assumption}{Assumption}
\newtheorem{theorem}{Theorem}

\theoremstyle{definition}

\newtheorem{exper}{Experiment}

\usepackage[sorting=none, giveninits=true, style=ieee]{biblatex}
\bibliography{references, extra}
\ifCLASSINFOpdf
\else
   \usepackage[dvips]{graphicx}
\fi
\usepackage{url}

\usepackage{subcaption}
\usepackage{graphicx}

\begin{document}

\title{Adaptively Optimised \\ Adaptive Importance Samplers}

\author[1]{Carlos A. C. C. Perello}
\author[1]{Deniz Akyildiz}
\affil[1]{Department of Mathematics, Imperial College London}

\maketitle

\begin{abstract}
We introduce a new class of adaptive importance samplers leveraging adaptive optimisation tools, which we term {AdaOAIS}. We build on \textit{optimised adaptive importance samplers} (OAIS), a class of techniques that adapt proposals to improve the mean-squared error of the importance sampling estimators by parameterising the proposal and optimising $\chi^2$-divergence between the target and the proposal. We show that a naive implementation of OAIS using stochastic gradient descent may lead to unstable estimators despite its convergence guarantees. To remedy this shortcoming, we instead propose to use adaptive optimisers (such as AdaGrad and Adam) to improve the stability of the OAIS. We provide convergence results for AdaOAIS in a similar manner to OAIS. We also provide empirical demonstration on a variety of examples and show that AdaOAIS lead to stable importance sampling estimators in practice.
\end{abstract}


\IEEEpeerreviewmaketitle

\section{Introduction}
\IEEEPARstart{A}{d}aptive importance samplers are a fundamental Monte Carlo method for estimating intractable integrals \cite{Robert2005MonteMethods}. In recent years, Adaptive Importance Sampling (AIS) techniques have seen wide use in many of science \cite{Wierling2012PredictionTreatment, Wang2020Three-dimensionalAirspace, He2022SensitivityCarlo}. At the heart of these problems lies the task of computing intractable expectations, which oftentimes are with respect to an unknown probability measure $\pi$. AIS schemes are an extension of classical importance sampling (IS) estimators, which uses a proposal distribution $q$ to estimate an expectation of the form $\mathbb{E}_\pi[\phi(X)]$, where $\phi$ is a bounded function and $\pi$ is an unknown target distribution. In AIS, this proposal is \textit{adapted} over iterations to improve performance. Some popular variants of AIS include population Monte Carlo methods \cite{Cappe2004PopulationCarlo, elvira2017improving, elvira2022optimized}, adaptive mixture importance sampling \cite{Cappe2007AdaptiveClasses}, layered AIS \cite{Martino2015LayeredSampling,martino2017anti,mousavi2021hamiltonian}, gradient-based AIS \cite{elvira2023gradient}, parametric AIS \cite{Akyildiz2021ConvergenceSamplers, Akyildiz2022GlobalSamplers} and multiple IS \cite{elvira2015efficient, elvira2019generalized}. A comprehensive review of AIS methods can be found in \cite{Bugallo2017AdaptiveFuture}.

As extensions of importance samplers, AIS routines also enjoy a $\mathcal{O}(1/\sqrt{N})$ convergence rate in $L_2$ (MSE), where $N$ is the number of points used to construct the empirical measure at each iteration \cite{Agapiou2017ImportanceCost}. Nevertheless, it is of interest to study the convergence rate of AIS in terms of both $N$ and $T$, the number of AIS iterations, to quantify the effect of \textit{adapting} the proposals. While it remains difficult to achieve a convergence rate for general AIS schemes over iterations, this has been proved possible for \textit{parametric} AIS in which the proposal $q$ is parameterised, denoted $q_\theta$ \cite{Akyildiz2021ConvergenceSamplers, Akyildiz2022GlobalSamplers}. In particular, \cite{Akyildiz2021ConvergenceSamplers} proposed a class of samplers termed \textit{optimised adaptive importance samplers} (OAIS), which is based on the minimisation of a loss function in order to adapt the proposal. Specifically, the variance of the weight function is chosen as the loss function to minimise; this idea has been well-studied in the literature \cite{Arouna2003RobbinsMonroFinance, Arouna2004AdaptativeTechnique, Kawai2008AdaptiveApproximation, Lapeyre2010AProcedures, Ryu2014AdaptiveProgramming, Kawai2017AccelerationApproximation, Kawai2018OptimizingApproximation}. OAIS are a class of importance samplers in which the parameter $\theta$ is optimised to minimise a loss function which is essentially the $\chi^2$-divergence between the target and the proposal. This allowed \cite{Akyildiz2021ConvergenceSamplers} to provide explicit convergence rates in terms of $N$ and $T$ when Stochastic Gradient Descent (SGD) is employed for minimisation (SGD OAIS). \\

\noindent\textbf{Contribution.} We begin this paper by showing the sensitivity issues of SGD when using OAIS, even in simple settings. In particular, we show that the optimisation procedure is sensitive to the step-size and this can cause non-robustness in more complex settings. To resolve this problem, we introduce a novel class of AIS estimators, namely \textit{Adaptive Gradient} OAIS (AdaOAIS). This class of AIS estimators minimise a quantity related to $\chi^2$-divergence using adaptive optimisers, which improve the stability of the optimisation routine by using gradient momentum terms. We will focus on the cases where AdaGrad \cite{Duchi2011AdaptiveOptimization} and Adam \cite{Kingma2014Adam:Optimization} are used. Finally, the same experiments that illustrate the numerical instability of SGD in the OAIS setting are repeated using AdaOAIS, showcasing the improved stability of AdaOAIS methods.
\\

\noindent\textbf{Notation.} Let $[T]:=\{0,\,1,\,\dots, \,T\}$. We use $q_\theta(x)$ and $\pi(x)$ to refer to the proposal and target densities, respectively. We write $\Pi(x)$ when referring to an unnormalised version of $\pi(x)$, where the normalisation constant is $Z:=\int_X\Pi(x) \md x$. Throughout the paper, $\phi(x)$ will be used to denote a bounded test function; for such functions, $(\phi,\,\mu) := \E_\mu[\phi(X)]$. The notation $\delta_z(\md x)$ denotes the Dirac measure centred at $z$. The parameter space of all possible values of $\theta$ will be denoted as $\Theta$. For two distributions $\mu,\,\nu$ such that $\mu,\,\nu \ll \lambda$, where $\lambda$ is the Lebesgue measure, let $D_{\chi^2}(\mu\|\nu)$ be their $\chi^2$ divergence.
\section{Technical Background}
We begin by introducing the Self-Normalised Importance Sampling (SNIS) estimator \cite{Robert2005MonteMethods}. For a given target and proposal distributions $\pi$ and $q_\theta$ respectively, the SNIS estimator is defined as
\begin{align}
(\phi, \pi) \approx (\phi, \pi_\theta^N) = \sum_{i=1}^N \mathsf{w}_\theta^{(i)} \phi(x_{\theta}^{(i)}),
\end{align}
where $x^{(i)}\sim q_\theta$ i.i.d., 
$\mathsf{w}_\theta^{(i)} = {W_\theta(x^{(i)})}/{\sum_{j=1}^N W_\theta(x^{(j)})}$ are normalised weights where $W_\theta(x) = \Pi(x)/q_\theta(x)$. Finally, $\pi_\theta^N$ is given by
$$
\pi^N_{\theta}(\md x) = \sum_{i=1}^N \mathsf{w}_\theta^{(i)} \delta_{x^{(i)}}(\mathrm{d} x).
$$
It is shown in \cite{Agapiou2017ImportanceCost} that the SNIS estimator $(\phi, \pi_\theta^N)$ enjoys the following convergence rate.
\begin{lemma}[{\cite[Theorem~2.1]{Agapiou2017ImportanceCost}}]
\label{thelemma}
If $(W_\theta^2,\,q_\theta)<\infty$, we have: 
\begin{align*}
\E[|(\phi,\,\pi)-(\phi,\,\pi^N_\theta)|^2]\leq\frac{4\|\phi\|^2_\infty\rho(\theta)}{N},
\end{align*}
where $\rho(\theta) := \E_{q_\theta}\left[\frac{\pi^2(X)}{q^2_\theta(X)}\right] = D_{\chi^2}(\pi\|q_\theta)+1$.
\end{lemma}
This bound, combined with the observation that $\rho(\theta)$ is convex when $q_\theta$ is an exponential family, see, e.g., \cite[Theorem~1]{Ryu2014AdaptiveProgramming} or \cite[Lemma~1]{Akyildiz2021ConvergenceSamplers}, led to a family of algorithms called optimised adaptive importance samplers (OAIS) \cite{Akyildiz2021ConvergenceSamplers,Akyildiz2022GlobalSamplers}. These methods aim at optimising $\rho(\theta)$ which also coincides with minimising the variance of the importance weights. OAIS methods are in contrast to population-based methods as they explicitly parameterise the proposal. The general method is given in Algorithm~\ref{alg:ais}.

\begin{algorithm}[t]
\caption{General OAIS algorithm \cite{Akyildiz2021ConvergenceSamplers}}
\label{alg:ais}
\begin{algorithmic}[1]
\State \text{Choose a proposal $q_\theta$ with initial parameter $\theta_0$}, a number of particles $N$.
\For{$k\geq 0$}
\State Sample $(x^{(i)}_k)_{i=1}^N\sim q_{\theta_k}$
\State Construct ${\pi}_{\theta_k}^N(\md x) = \sum_{i=1}^N \mathsf{w}_\theta^{(i)} \delta_{x^{(i)}_k}(\md x)$
\State Report $(\phi,\,{\pi}^N_{\theta_k})$
\State Compute the updated parameter $\theta_{k+1}=\mathcal{T}(\theta_k)$
\EndFor
\end{algorithmic}
\end{algorithm}

However, we cannot typically compute $\rho(\theta)$, but instead some unnormalised version $R(\theta) := \E_{q_{\theta}}\left[{\Pi^2(X)}/{q^2_\theta(X)}\right] = Z^2\rho(\theta)$. In OAIS, the update rule $\mathcal{T}:\Theta\to\Theta$ is defined using the gradients of $R(\theta)$, such as stochastic gradient descent or, in our case, adaptive optimisers. Therefore, it is of interest to find convergence rates of the different OAIS algorithms in terms of $N$ and $T$. In \cite{Akyildiz2021ConvergenceSamplers}, SGD OAIS was analysed and, under mild assumptions, was found to have rate $\mathcal{O}(1/N + {\log T}/{N\sqrt{T}})$. In order to streamline the discussion about the convergence of OAIS schemes, we introduce the notion of \textit{adaptive rate}; if an OAIS algorithm has rate $\mathcal{O}(f(T)/N + 1/N)$ where $f(T)\to0$ as $T\to \infty$, we say that the OAIS algorithm has adaptive rate $\mathcal{O}_a(f(T))$. If $f(T)\not\to 0$ as $T\to\infty$, we will say that the OAIS algorithm in question has no adaptive rate. Using this framework, SGD OAIS obtains an adaptive rate of $\mathcal{O}_a(\log T/{\sqrt{T}})$. Nevertheless, we remark that SGD OAIS can be sensitive to its parameters like the step-size. To give a simple example, consider $\pi\sim\mathcal{N}(\mu_\pi,\,\sigma^2)$ and $q_{\theta}\sim\mathcal{N}(\theta,\,\sigma^2)$ are both one-dimensional Gaussians with equal variance. In this case, we have $\rho(\theta)\propto \exp((\mu_\pi- \theta)^2)$. This loss function and its gradients may become explosive depending on the step-size.

\begin{figure}[t]
\centering\includegraphics[scale=0.32]{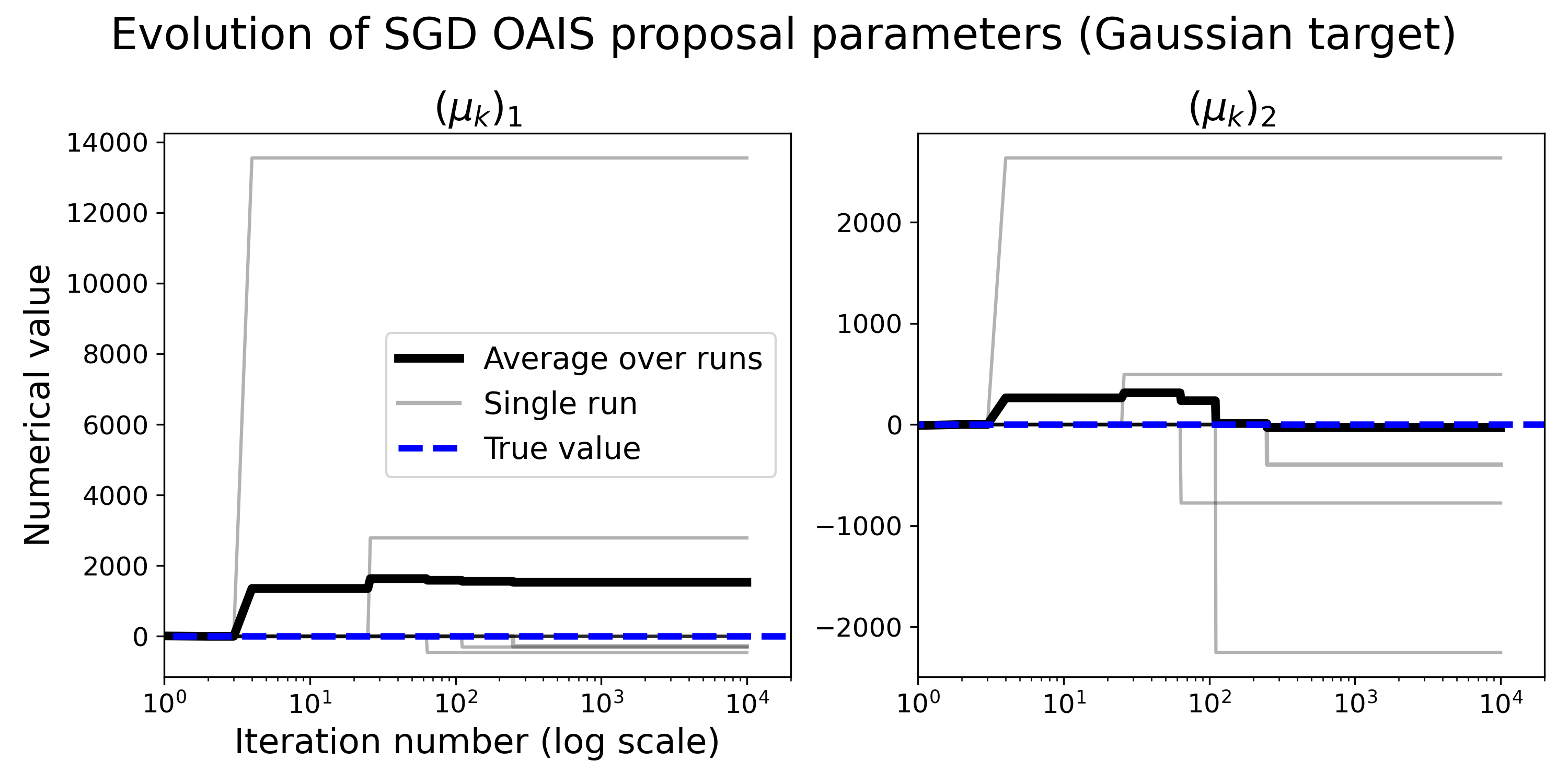}
    \caption{The divergence of SGD with a decreasing step-size $t_k={10^{-4}}/{\sqrt{k+1}}$. We plot here the entries of $\mu_k$ over time when using SGD OAIS. Covariance entries exhibit similar behaviour.}
    \label{fig:sgd-fail}
\end{figure}

A more general example on 2D Gaussians is demonstrated in Figure \ref{fig:sgd-fail} where the optimisation process diverges. This motivates the introduction of AdaOAIS using adaptive optimisers, which are typically much more stable than SGD in terms of sensitivity.

To arrive at AdaOAIS, an empirically faster and more numerically stable class of OAIS algorithms, we first introduce adaptive optimisers. Adaptive optimisers are a type of optimisers that keep track of previous gradient estimates to prevent an anomalous gradient estimate from destabilising the optimisation routine. Formally, this is achieved by computing regular or heavy-ball \textit{momentum} terms at each iteration \cite{Polyak1964SomeMethods}. These momentum terms are then used to smoothen out any effects an anomalous gradient estimate may incur on the procedure. The two most popular adaptive optimisers are Adam \cite{Kingma2014Adam:Optimization}, which employs both regular and heavy-ball terms, and AdaGrad \cite{Duchi2011AdaptiveOptimization}, which only uses regular momentum. 
\section{Adaptively Optimised AIS (AdaOAIS)}
Adaptively Optimised AIS (AdaOAIS) techniques use adaptive optimisers to find the optimal proposal distribution. In this work, we mainly focus on the scenarios where Adam and AdaGrad are utilized to fine-tune the proposal's parameter, $\theta$. We call Adam OAIS and AdaGrad OAIS the OAIS algorithms that use Adam and AdaGrad to update $\theta$, respectively. We now give the parameter update rules for the AdaOAIS algorithms in terms of $\mathcal{T}(\theta_k)$, following the notation used in Algorithm \ref{alg:ais}. Let $g_k := g(\theta_k)$.\\

\noindent\textbf{Adam OAIS.} Adam OAIS method uses Adam optimiser to update the parameter. In other words, we define $\mathcal{T}_{\text{Adam}}(\theta_k)$
\begin{align*}
    \mathcal{T}_{\text{Adam}}(\theta_k) &= \theta_k - t_k \odot ({\hat{m}_{k+1}}\oslash({\sqrt{\hat{v}_{k+1}}+\varepsilon))
    }
    \end{align*}
    where $\forall k\geq 0$:
\begin{align*}
    m_{k+1} &= \beta_1 \odot m_k + (1-\beta_1)\odot g_k\\
    \hat{m}_{k+1} &= {m_{k+1}}\oslash({1-\beta_1^{k+1}})\\
    v_{k+1} &= \beta_2 \odot v_k + (1-\beta_2)\odot g_k^{\odot^2}\\
    \hat{v}_{k+1} &= {v_{k+1}}\oslash({1-\beta_2^{k+1}})
\end{align*}
and $m_0,\,v_0 = 0$. The parameter $\varepsilon$ is used to avoid any numerical errors; in Adam OAIS we take $\varepsilon=10^{-8}$. \\

\noindent\textbf{AdaGrad OAIS.} Similarly, the update rule for AdaGrad OAIS is defined using the AdaGrad optimiser:
    \begin{align*}
    \mathcal{T}_{\text{AdaGrad}}(\theta_k) &= \theta_k - t_k \odot ({g_{k}}\oslash(\text{diag}({\sqrt{G_k} + \varepsilon I)})),
\end{align*}
where $\forall k \geq 0$ and
\begin{align*}
    G_k &= \sum_{n=1}^k g_ng_n^\top.
\end{align*}
Once more, we take $\varepsilon=10^{-8}$ for numerical stability.
\section{Convergence rates}
We introduce the necessary assumptions first.
\begin{assumption}
\label{strongconvass}
    $R(\theta)$ is $\mu$-strongly convex and $L$-smooth.
\end{assumption}

\begin{figure*}[t!]
\centering
\includegraphics[scale=0.35]{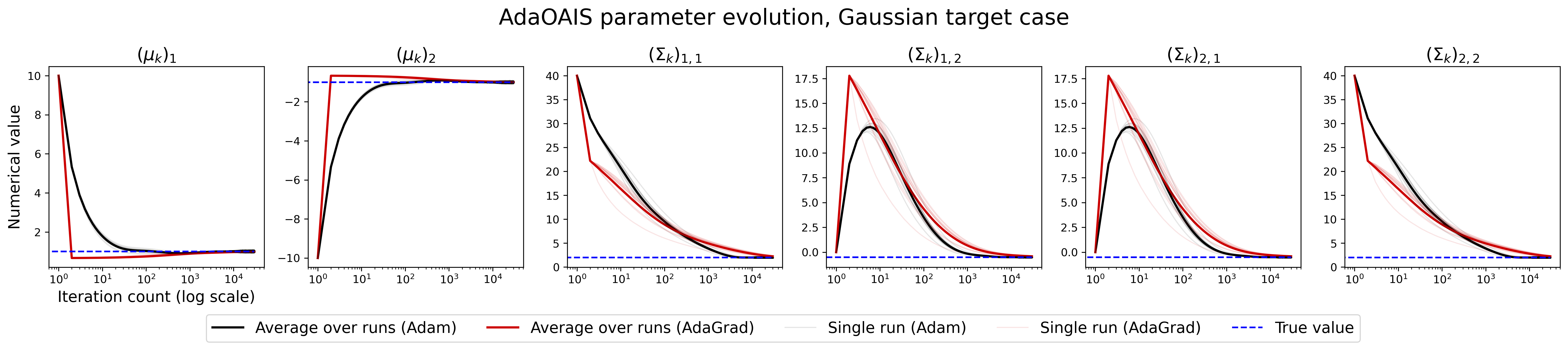}
\caption[Gaussian Adam OAIS]{Evolution of the parameters in Experiment~\ref{exp:Gaussian_target} (Gaussian target). AdaOAIS entry-wise convergence of $(\mu_k,\,\Sigma_k)$ to $(\mu_\pi,\,\Sigma_\pi)$ can be seen in all entries of the mean vector and the covariance matrix. We see numerically stable convergence to the true parameter values.}\label{fig:adagradparams}
\end{figure*}
Assumption~\ref{strongconvass} is motivated by the fact that $R$ is in general convex when $q_\theta$ belongs to the exponential family \cite{Ryu2014AdaptiveProgramming, Akyildiz2021ConvergenceSamplers}. In certain cases, $\mu$-strong convexity holds (e.g. for two Gaussians with equal variance). For ease of analysis, we keep the strong-convexity assumption, as the analysis below can be adapted for the convex case.

\begin{assumption}
\label{unbiasedg}
    $g(\theta)$ is an unbiased estimator of $\nabla R(\theta)$ and is almost surely bounded, i.e. $\exists K,\,\varepsilon>0$ such that:
    $$
    \|g(\theta)\|_\infty \leq K-\sqrt{\varepsilon} \quad \text{a.s.}
    $$
\end{assumption}

Using these assumptions, we are now prepared to give our first result.

\begin{theorem} (Adam OAIS)
\label{AdamCR}
     Suppose that Assumption \ref{strongconvass} holds for $R(\theta)$ and that Assumption \ref{unbiasedg} holds for $g(\theta)$. Let $\theta_k$ be the iterates generated by running Adam OAIS with hyperparameters $t_k = \alpha,\,\beta_2\in(0,\,1),\,\beta_1\in(0,\,\beta_2)$ and $\varepsilon> 0$ for $T\geq k$ iterations starting at $\theta_0$, where $T\geq {\beta_1}/({1-\beta_1})$. Let $(\theta_k)_{k\geq 1}$ be the iterates generated after $k$ iterations of Adam OAIS Then the following bound holds: 
     \begin{gather*}
         \min_{k\in[T]}\E\left[|(\phi,\,\pi) - (\phi,\,{\pi}^N_{\theta_{k}})|^2\right] \leq \frac{4K\|\phi\|_\infty^2}{\alpha\mu Z^2} \frac{R(\theta_0)-R(\theta^\star)}{N(\tilde{T}+1)} \\+ \frac{4E\|\phi\|_\infty^2}{N(\tilde{T}+1)Z^2}\left[\log\left(1+\frac{K^2}{(1-\beta_2)\varepsilon}\right)-(T+1)\log(\beta_2)\right] \\+ \frac{4\|\phi\|_\infty^2R(\theta^\star)}{Z^2N}
    \end{gather*}
where $E=E(R,\,L,\,d,\,t,\,\beta_1,\,\beta_2,\,\mu)$ and $\tilde{T} = T-\frac{\beta_1}{1-\beta_1}$.
\end{theorem}
\begin{proof}
    See Appendix.
\end{proof}
We see that the rate of Adam OAIS is $\mathcal{O}(1/N + f_1(T)/N)$. However, note that as $T\to \infty$, $f_1(T)\not\to 0$ due to the $(T+1)$ term present in front of $\log\beta_2$; this implies that Adam OAIS does not have an adaptive rate. Furthermore, if we let $T\to\infty$ we recover another term in addition to the term present in Theorem \ref{thelemma}. Next, we also provide rates for AdaGrad OAIS below which does not suffer from this problem.
    \begin{theorem} (AdaGrad OAIS)
    \label{AdaGradCR}
    Suppose that Assumption \ref{strongconvass} holds for $R(\theta)$ and that Assumption \ref{unbiasedg} holds for $g(\theta)$. Let $\theta_k$ be the iterates generated by running AdaGrad OAIS with constant learning rate $t_k=\alpha$. Then the following bound holds:
    \begin{gather*}
     \min_{k\in[T]}\E\left[|(\phi,\,\pi) - (\phi,\,{\pi}^N_{\theta_{k}})|^2\right] \leq \frac{4\|\phi\|_\infty^2}{\mu\alpha Z^2}\frac{R(\theta_0) - R(\theta^\star)}{N\sqrt{T+1}} \\+ \frac{(8dK^2 + 2\alpha dKL)\|\phi\|_\infty^2}{\mu Z^2 N\sqrt{T+1}}\log\left(1+\frac{(T+1)K^2}{\varepsilon}\right) \\+ \frac{4\|\phi\|_\infty^2R(\theta^\star)}{Z^2N}
    \end{gather*}
\end{theorem}
\begin{proof}
    See Appendix.
\end{proof}
In this case, writing the convergence rate of AdaGrad OAIS as $\mathcal{O}(1/N + f_2(T)/N)$ results in $f_2(T)\to0$ as $T\to\infty$. Therefore, switching from Adam OAIS to AdaGrad OAIS allows us to recover an adaptive convergence rate, namely $\mathcal{O}_a(\log T/{\sqrt{T}})$. Observe that if in Theorem \ref{AdamCR} the $\log\beta_2$ were removed, one would obtain a faster adaptive rate of $\mathcal{O}_a(1/T)$. This is as expected as it is known that, in general, Adam converges to the minimum faster than AdaGrad.


\section{Experimental results}

We now display some numerical results to empirically verify the bounds derived for the AdaOAIS algorithms.
\begin{exper}\label{exp:Gaussian_target} Let $\pi(x) = \mathcal{N}(x;\mu_\pi,\,\Sigma_\pi)$ where: 
$$\mu_\pi = \begin{bmatrix}
    1\\-1
\end{bmatrix},\quad\Sigma_\pi = \begin{bmatrix}
    2 & -0.5\\ -0.5 & 2
\end{bmatrix}.$$
We call this case the \textit{Gaussian target} case. In this setting, the full parameter space is $\Theta = \R^2\times \text{PD}^2$, where $\text{PD}^2$ is the cone of $2\times 2$ positive-definite matrices. We estimate $\E_\pi[1_D(X)] = \mathbb{P}(X \in D)$, where $D = [-1,\,1]\times [-1,\,1]$. To achieve this, we use Normally distributed proposals -- let $q_k(x) := q_{\theta_k}(x) = \mathcal{N}(x; \mu_k,\,\Sigma_k)$. We shall start with a Gaussian proposal $q_0(x) = \mathcal{N}(x; \mu_0,\,\Sigma_0)$, where:
$$\mu_0 = \begin{bmatrix}
    10\\-10
\end{bmatrix},\quad\Sigma_0 = \begin{bmatrix}
    40 & 0\\ 0 & 40
\end{bmatrix}.$$
\begin{figure}[t!]
    \centering
    \begin{subfigure}[t]{0.5\columnwidth}
        \centering
\includegraphics[scale=0.22]{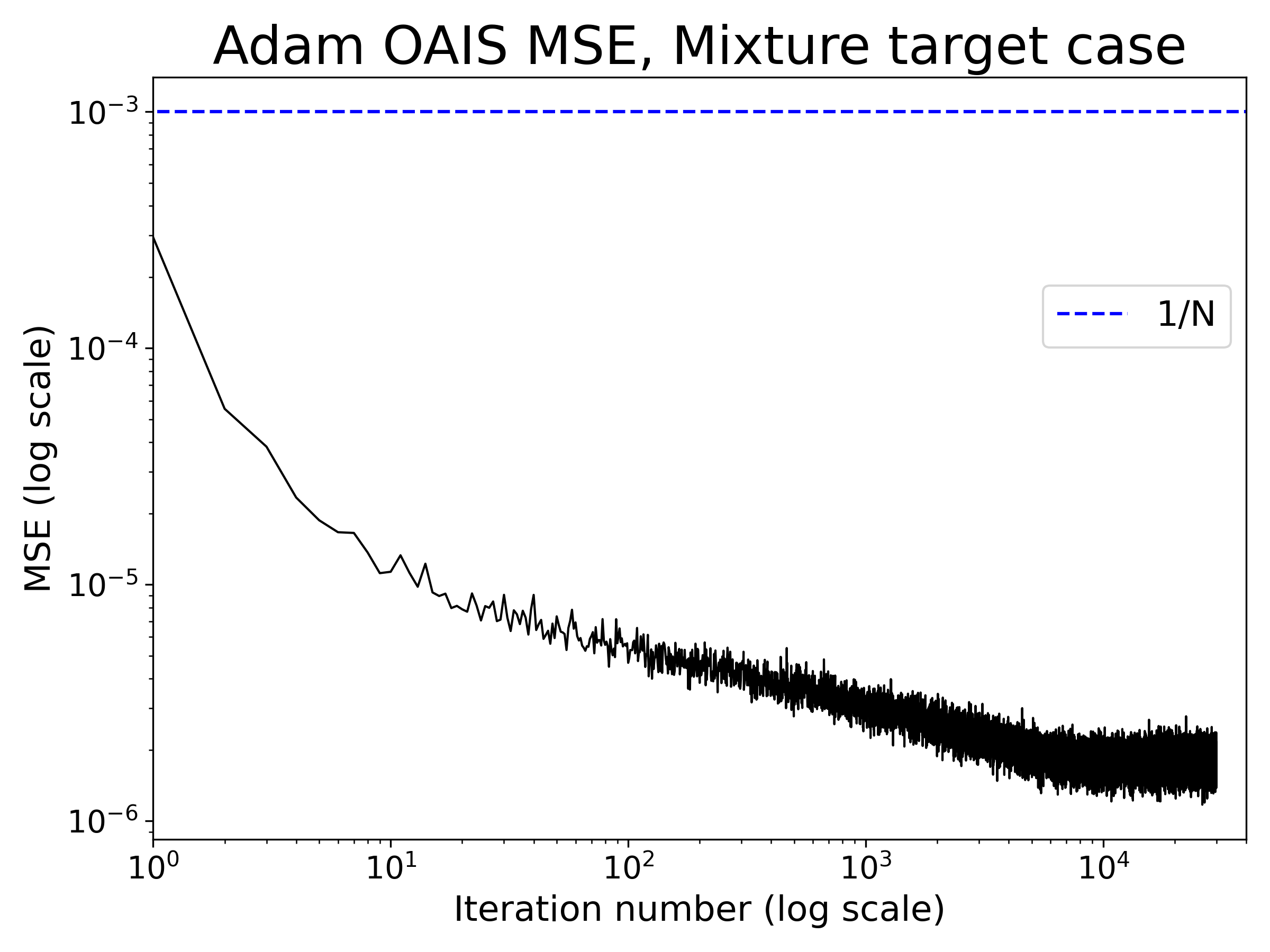}
    \caption{}\label{fig:mse_adam_mix}
    \end{subfigure}%
    \begin{subfigure}[t]{0.5\columnwidth}
        \centering
\includegraphics[scale=0.22]{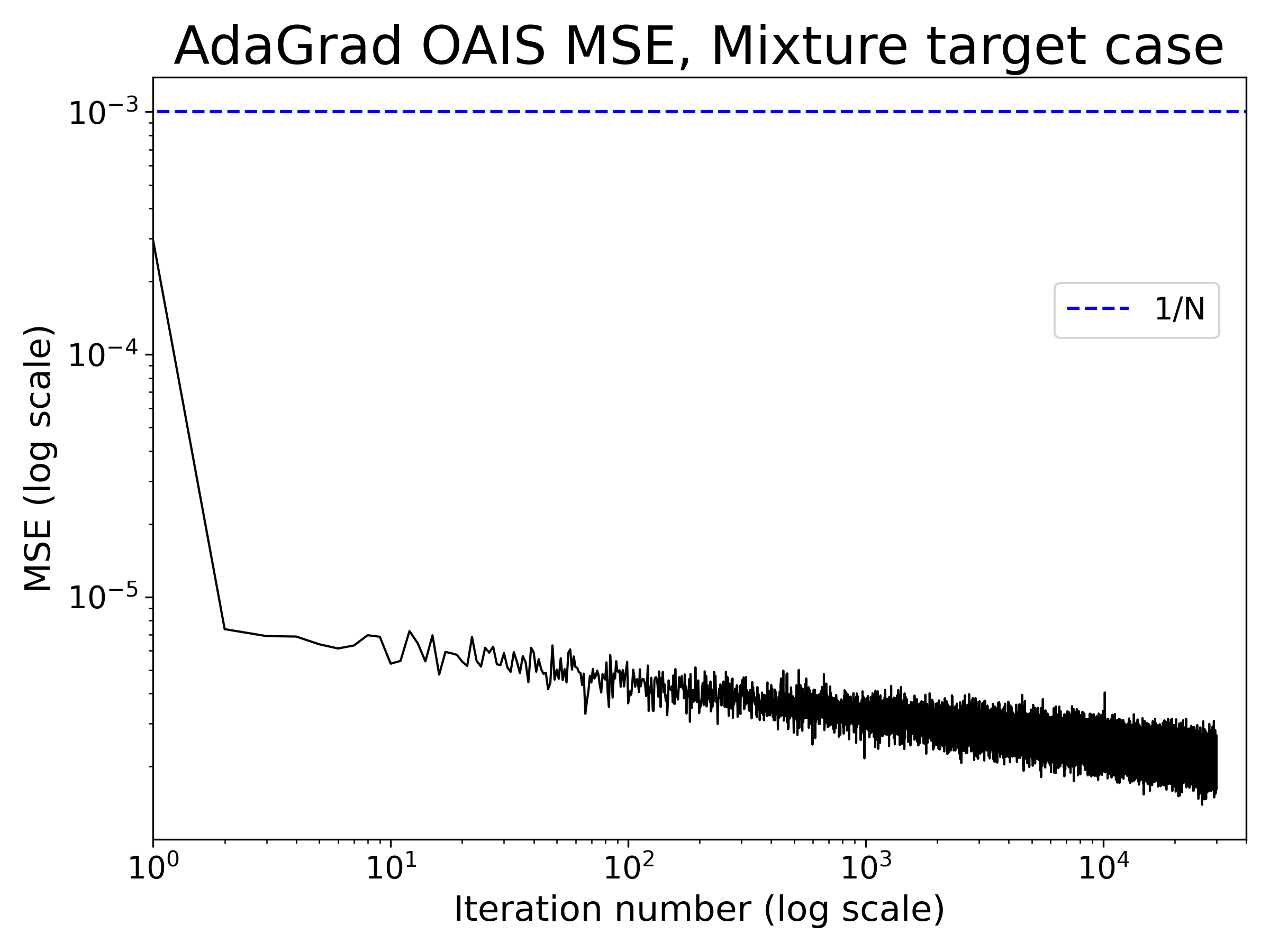}
\caption{}\label{fig:mse_adagrad_mix}
    \end{subfigure}
    \caption{The MSEs over iterations in Experiment~\ref{exp:mixture_of_target} (Mixture target). A similar trend is observed in the Gausian target case. (a) Adam OAIS. (b) AdaGrad OAIS.}\label{fig:mse}
\end{figure}
For both algorithms, $N=1000$ particles were used to construct the empirical measure at each iteration and $10$ runs were performed to compute the MSE. Initially, SGD OAIS was run with a learning rate of $t_k = {10^{-4}}/{\sqrt{k+1}}$ for $T=10000$ iterations. Figure \ref{fig:sgd-fail} shows the entries of $\mu_k$ as the iteration number increases. It can be easily seen that even with a small initial learning rate, SGD OAIS might be numerically unstable.

Subsequently, Adam OAIS and AdaGrad OAIS were executed. Adam OAIS was run as in Theorem \ref{AdamCR}, with $t_k = 0.01,\,\beta_1=0.9,\,\beta_2=0.999$ for $T=30000$ iterations, whilst AdaGrad OAIS was also run as in Theorem \ref{AdaGradCR} with $t_k=0.1$ for a total of $T=30000$ iterations.

Figure \ref{fig:adagradparams} shows the evolution of the parameter entries for Adam and AdaGrad OAIS. In these settings, it can be seen that the parameters evolve in a much more numerically stable manner, as the stray grey lines observed in the SGD OAIS case are no longer present. Furthermore, both $\mu_k$ and $\Sigma_k$ converge entry-wise to $\mu_\pi$ and $\Sigma_\pi$. Finally, observe that Adam OAIS converges in a smoother way than AdaGrad OAIS -- this is due to the inclusion of heavy-ball momentum in Adam OAIS, which smoothens the optimisation routine. \end{exper}
\begin{exper}\label{exp:mixture_of_target} We now consider a mixture target $\pi$ where a Gaussian proposal density initialised at $q_0$ was used to estimate $\mathbb{P}(X \in D)$, where $X\sim\pi$, $q_0$ and $D$ are as in the Gaussian target case. The target considered is a bimodal Gaussian mixture with equal weights:
\begin{equation*}
    \pi(x) = w_1 \mathcal{N}(x;m_1,\Sigma_1) + w_2 \mathcal{N}(x; m_2, \Sigma_2),
\end{equation*}
where $m_1= \begin{bmatrix}
    3& 0
\end{bmatrix}^\top$, $m_1= \begin{bmatrix}
    -3& 0
\end{bmatrix}^\top,\,\Sigma_1 = \Sigma_2 = I_2$ and $w_1=w_2=0.5$. We refer to this scenario as the \textit{Mixture target} case. In this scenario, both Adam OAIS and AdaGrad OAIS were executed with a fixed learning rate of $t_k=0.01$ and $t_k = 0.1$, respectively, for a total of $T=30000$ iterations per run. Adam OAIS used the hyperparameters $\beta_1 = 0.9$ and $\beta_2=0.999$. To compute the MSE, $200$ runs of the AdaOAIS algorithms were performed. Figures \ref{fig:mse_adam_mix} and \ref{fig:mse_adagrad_mix} show the MSE of AdamOAIS and AdamOAIS in the Mixture target case as a function of the iteration number. For all cases, the MSE is below $1/N$ for all iterations. As $1/N$ is a tighter bound than the one presented in Theorems \ref{AdamCR} and \ref{AdaGradCR}, Figures \ref{fig:mse_adam_mix} and \ref{fig:mse_adagrad_mix} empirically verify Theorems \ref{AdamCR} and \ref{AdaGradCR}.
\end{exper}

\begin{exper}\label{exp:logit_normal} Finally, we consider a nontrivial scenario where AdaOAIS proves particularly useful. We aim at estimating $\mathbb{P}(X \in D)$, where $X \sim \text{LogitNormal}(x; 0,\,1) = \pi(x)$ and $D=[0.25,\,0.75]$. The Logit Normal distribution is a highly intractable distribution, and only recently analytic (but not closed) form of its moments \cite{Holmes2022MomentsDistribution} were derived. Crucially, the Logit Normal distribution does \textit{not} belong to the exponential family of distributions. We use $q_\theta(x) = \text{Beta}(x;\alpha,\,\beta)$ where $\theta = (\alpha,\beta)$, as the Logit Normal distribution is supported in $(0,\,1)$ and the Beta distribution belongs to the exponential family of distributions, guaranteeing the existence of an optimal parameter $\theta^\star$. We choose $\theta_0 = (1, 1)$ and note that the parameter space is the cone $\Theta = \R_{>0}\times\R_{>0}$.

\begin{figure}[t!]
    \centering
    \includegraphics[scale=0.17]{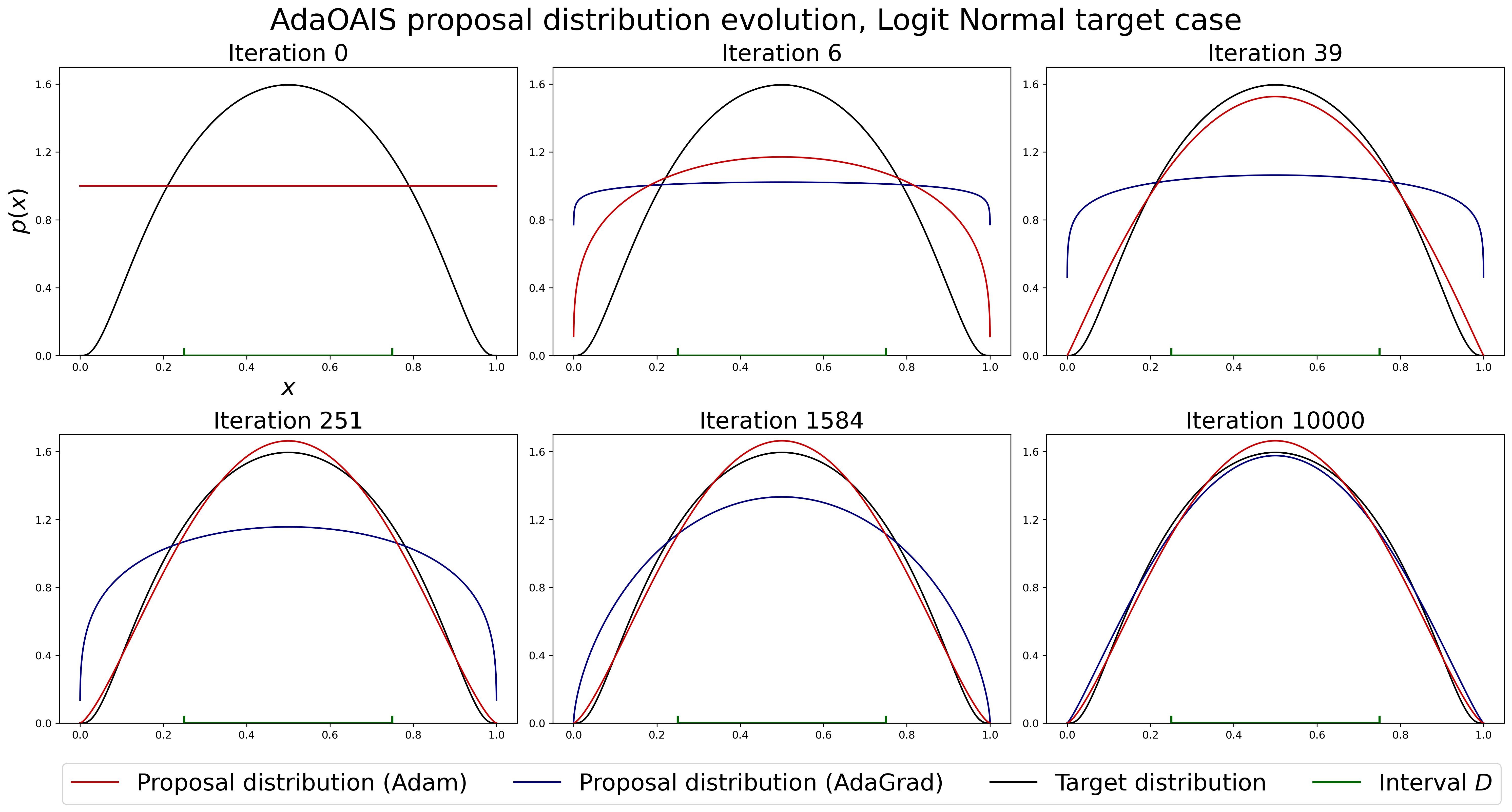}
    \caption{The evolution of the proposals in Experiment~\ref{exp:logit_normal} (Logit-Normal target). The proposals were averaged over $100$ runs.}
    \label{fig:beta_proposal}
\end{figure}

Once more, we empirically verify the efficacy of the AdaOAIS algorithms. We executed both Adam OAIS and AdaGrad OAIS for a total of $T=10000$ iterations over $100$ runs, using a learning rate of $t_k = 0.01$ for Adam OAIS and $t_k = 0.1$ for AdaGrad OAIS. Figure \ref{fig:beta_proposal} shows the change in the proposals (averaged over all runs) at logarithmically-spaced iteration numbers. It can be seen that Adam OAIS proposals converge faster than the AdaGrad OAIS proposals to the target distribution.
\end{exper}
\section{Conclusion}
In this letter, we introduced a novel family of Adaptive Importance Sampling (AIS) techniques, AdaOAIS. We proved convergence rates for AdaOAIS and provided numerical examples which motivate the use of AdaOAIS and verified their theoretical bounds. Future works include providing similar results non-convex (general) $R(\theta)$ \cite{Akyildiz2022GlobalSamplers} or the use of similar adaptation steps within a particle filter to improve performance similar to nudging steps, e.g. \cite{akyildiz2020nudging}.

\newpage

\printbibliography
\newpage
\appendix
\label{theappendix}
\section{Proofs}

\begin{proof}[Proof of Theorem 1]

Let $\mathcal{F}^A_n$ be the filtration generated by running Adam OAIS for $n$ steps. Using Lemma \ref{thelemma} and taking conditional expectations with respect to $\mathcal{F}^A_{k-1}$ gives:
    \begin{align*}
        \E\left[|(\phi,\,\pi) - (\phi,\,{\pi}^N_{\theta_{k}})|^2|\mathcal{F}^A_{k-1}\right] &= |(\phi,\,\pi) - (\phi,\,{\pi}^N_{\theta_{k}})|^2\\&\leq \frac{4\|\phi\|_\infty^2 R(\theta_k)}{Z^2N}
    \end{align*}
    Adding and subtracting $\frac{4\|\phi\|_\infty^2R(\theta^*)}{Z^2N}$ to the RHS we obtain and taking expectations once more yields:
\begin{align}
         &\min_{k\in[T]}\E\left[|(\phi,\,\pi) - (\phi,\,{\pi}^N_{\theta_{k}})|^2\right]\nonumber \\&\leq \frac{4\|\phi\|_\infty^2}{Z^2N}\min_{k\in[T]}\E[R(\theta_k)-R(\theta^*)] + \frac{4\|\phi\|_\infty^2R(\theta^*)}{Z^2N}\label{adamineq}
\end{align}
Finally, observe that as $R(\theta)$ is $\mu$-strongly convex by Assumption \ref{strongconvass}, it satisfies the Polyak-Łojasiewicz inequality \cite{Polyak1963GradientFunctionals}, and thus $\forall n\in [T]$:
\begin{equation*}
    \min_{k\in[T]}\E[R(\theta_k)-R(\theta^*)]\leq \frac{1}{2\mu} \E[\|\nabla R(\theta_n)\|_\Theta^2]
\end{equation*}
By  Assumption \ref{unbiasedg}, one can apply Theorem 4 of \cite{Defossez2020AAdagrad} together with the inequality above to yield:
\begin{align*}
    &\min_{k\in[T]}\E[R(\theta_k)-R(\theta^*)]\leq  \frac{K}{\alpha\mu} \frac{R(\theta_0)-R(\theta^*)}{{T}+1} \\&+ E'\left(\frac{1}{{T}+1}\log\left(1+\frac{K^2}{(1-\beta_2)\varepsilon}\right)-\frac{T+1}{{T}+1}\log(\beta_2)\right)
\end{align*}
Where $E$ is a constant depending only on $K,\,L,\,d,\,t,\,\beta_1,\,\beta_2$ and $\mu$ and ${T} = T-\frac{\beta_1}{1-\beta_1}$.

Plugging this into \eqref{adamineq} gives the desired inequality:
\begin{gather*}
         \min_{k\in[T]}\E\left[|(\phi,\,\pi) - (\phi,\,{\pi}^N_{\theta_{k}})|^2\right] \leq \frac{4K\|\phi\|_\infty^2}{\alpha\mu Z^2} \frac{R(\theta_0)-R(\theta^*)}{N({T}+1)} \\+ \frac{4E\|\phi\|_\infty^2}{N({T}+1)Z^2}\left[\log\left(1+\frac{K^2}{(1-\beta_2)\varepsilon}\right)-(T+1)\log(\beta_2)\right] \\+ \frac{4\|\phi\|_\infty^2R(\theta^*)}{Z^2N}
    \end{gather*}
\end{proof}

\begin{proof}[Proof of Theorem 2]

Let $\mathcal{F}^{AG}_n$ be the filtration generated by running AdaGrad OAIS for $n$ steps. Using Lemma \ref{thelemma} and taking conditional expectations with respect to $\mathcal{F}^A_{k-1}$ gives:
    \begin{align*}
        \E\left[|(\phi,\,\pi) - (\phi,\,{\pi}^N_{\theta_{k}})|^2|\mathcal{F}^{AG}_{k-1}\right] &= |(\phi,\,\pi) - (\phi,\,{\pi}^N_{\theta_{k}})|^2\\&\leq \frac{4\|\phi\|_\infty^2 R(\theta_k)}{Z^2N}
    \end{align*}
    Once more, adding and subtracting $\frac{4\|\phi\|_\infty^2R(\theta^*)}{Z^2N}$ to the RHS we obtain and taking expectations gives:
\begin{align}
         &\min_{k\in[T]}\E\left[|(\phi,\,\pi) - (\phi,\,{\pi}^N_{\theta_{k}})|^2\right]\nonumber \\&\leq \frac{4\|\phi\|_\infty^2}{Z^2N}\min_{k\in[T]}\E[R(\theta_k)-R(\theta^*)] + \frac{4\|\phi\|_\infty^2R(\theta^*)}{Z^2N}\label{adagradineq}
\end{align}
Again, as $R(\theta)$ is $\mu$-strongly convex by Assumption \ref{strongconvass}, it also the Polyak-Łojasiewicz inequality \cite{Polyak1963GradientFunctionals}, and thus $\forall n\in [T]$:
\begin{equation*}
    \min_{k\in[T]}\E[R(\theta_k)-R(\theta^*)]\leq \frac{1}{2\mu} \E[\|\nabla R(\theta_n)\|_\Theta^2]
\end{equation*}
By  Assumption \ref{unbiasedg}, one can apply Theorem 1 of \cite{Defossez2020AAdagrad} together with the inequality above to yield:
\begin{align*}
&\min_{k\in[T]}\E[R(\theta_k)-R(\theta^*)]\leq K\frac{R(\theta_0) - R(\theta^*)}{\mu\alpha \sqrt{T+1}} \\&+ \frac{4dK^2 + \alpha dKL}{2\mu\sqrt{T+1}}\log\left(1+\frac{(T+1)K^2}{\varepsilon}\right)
\end{align*}
Finally, plugging the above inequality into \eqref{adagradineq} yields the final result:
\begin{gather*}
     \min_{k\in[T]}\E\left[|(\phi,\,\pi) - (\phi,\,{\pi}^N_{\theta_{k}})|^2\right] \leq \frac{4\|\phi\|_\infty^2}{\mu\alpha Z^2}\frac{R(\theta_0) - R(\theta^*)}{N\sqrt{T+1}} \\+ \frac{(8dK^2 + 2\alpha dKL)\|\phi\|_\infty^2}{\mu Z^2 N\sqrt{T+1}}\log\left(1+\frac{(T+1)K^2}{\varepsilon}\right) \\+ \frac{4\|\phi\|_\infty^2R(\theta^*)}{Z^2N}
    \end{gather*}
\end{proof}

\end{document}